\documentclass[12pt]{article}
\usepackage[english]{babel}
\usepackage[latin1]{inputenc}
\usepackage{fancyhdr,indentfirst,graphicx,newlfont,amssymb,amsmath,latexsym,amsthm}
\usepackage{subfigure}
\usepackage[latin1]{inputenc}
\usepackage{longtable}
\usepackage{color}
\begin{document}
\thispagestyle{empty}
\newtheorem{df}{Definition}
\newtheorem{prop}{Proposition}
\newtheorem{rem}{Remark}
\newtheorem{cor}{Corollary}
\newtheorem{es}{Example}
\newtheorem{teo}{Theorem}
\newtheorem{lm}{Lemma}
\newtheorem{pr}{Property}
\noindent A GENERALIZATION OF THE SKEW-NORMAL DISTRIBUTION: THE BETA SKEW-NORMAL
\vskip 3mm

\vskip 5mm
\noindent Valentina Mameli and Monica Musio

\noindent Dipartimento di Matematica

\noindent  Università di Cagliari

\noindent Via Ospedale 72 09124, Italy

\noindent vmameli@unica.it

\vskip 3mm

\begin{abstract}
We consider a new generalization of the skew-normal distribution introduced by Azzalini  (1985). We denote this distribution Beta skew-normal (BSN) since it is a special case of the Beta generated distribution (Jones (2004)). Some properties of the BSN are studied. We pay attention to some  generalizations of the skew-normal distribution (Bahrami et al. (2009), Sharafi and Behboodian (2008), Yadegari et al. (2008)) and to their relations with the BSN. 
\end{abstract}

\section{Introduction}
The skew-normal distribution (SN), introduced by Azzalini (1985), has been studied and generalized extensively. The aim of this article is to introduce a new family of distributions, which generalizes the skew normal, that is flexible enough to support both unimodal and bimodal shape. This new family, called Beta skew-normal (BSN), arises naturally when we consider the distributions of order statistics of the skew-normal distribution. The BSN can also be obtained as a special case of the Beta generated distribution (Jones (2004)). In this work we pay attention to three other generalizations of the skew-normal distribution: the Balakrishnan skew-normal (SNB) (Balakrishnan (2002), as a discussant of Arnold and Beaver (2002), Gupta and Gupta (2004), Sharafi and Behboodian (2008)), the generalized Balakrishnan skew-normal (GBSN) (Yadegari et al. (2008)) and a two parameter generalization of the Balakrishnan skew-normal (TBSN) (Bahrami et al. (2009)).
The above three extensions are related to the Beta skew-normal distribution for particular values of the parameters.\\
Given a random sample $X_{1},\cdots,X_{n}$ from a distribution $F(x)$, in general the distribution of the related order statistics does not belong to the family of $F(x)$. In this paper we show that the maximum between the $X_{i}$'s from a Balakrishnan skew-normal with parameters $m$ and $1$, denoted by $X_i \sim SNB_{m}(1)$, is still a Balakrishnan skew-normal with parameters $k$ and $1$, where $k$ is a function of $m$ and $n$. \\
This paper is organized as follows: after describing briefly, in Section 2, the skew-normal distribution, its generalizations  and listing their most important properties, in Section 3 we present some generalizations of the Beta distribution.
In the last Section we define the Beta skew-normal distribution, we present its properties and some special cases. In particular the BSN contains the Beta half-normal distribution (Pescim et al. (2010)) as limiting case. Besides, we investigate its shape properties. We derive its moment generating function and we also compute numerically the first moment, the variance, the skewness and the kurtosis. We present two different methods which allow to simulate a BSN distribution. We explore its relationships with the other generalizations of the skew-normal and we show that the distributions of order statistics from the skew-normal distribution are Beta skew-normal and are log-concave. Furthermore, in this Section we give some results concerning the SNB distribution. In particular we derive the exact distributions of the largest order statistic from $SNB_{m}(1)$ and the shortest order statistic from $SNB_{m}(-1)$.
\section{The skew-normal density and its generalizations}
The present Section recalls some important definitions and properties about the skew-normal distribution and some of its extensions.
\subsection{The skew-normal density}
The skew-normal distribution refers to a parametric class of probability distributions which includes the standard normal as a special case. A random variable $Z$ is said to be skew-normal with parameter $\lambda$, if its density function is
\begin{equation}
\phi(z;\lambda)=2\phi(z)\Phi(\lambda z)\quad\textrm{with}\quad \lambda, z\in\mathbb{R} \label{prima}
\end{equation}
where $\phi(\cdot)$ and $\Phi(\cdot)$ denote the standard normal density and distribution, respectively. We denote a random variable Z with the above density by $Z\sim SN(\lambda)$. The parameter $\lambda$ controls skewness. The standard normal distribution is a skew-normal distribution with $\lambda=0$.
We remind some properties of the SN distribution.\\
\textbf{Properties of $SN(\lambda)$}:
\begin{itemize}
\item[a.] As $\lambda\rightarrow\infty$, $\phi(z;\lambda)$ tends to the half-normal density.
\item[b.] If Z is a $SN(\lambda)$ random variable, then $-Z$ is a $SN(-\lambda)$ random variable.
\item[c.] If $Z\sim SN(\lambda)$, then $Z^{2}\sim \chi^{2}(1)$.
\item[d.] The density (\ref{prima})
is strongly unimodal, i.e. $\log{\phi(z;\lambda)}$ is a concave function of $z$.
\end{itemize}
The corresponding distribution function is $$\Phi(z;\lambda)=2\int_{-\infty}^{z}\int_{-\infty}^{\lambda t}\phi(t)\phi(u)du dt=\Phi(z)-2T(z,\lambda),$$ where $T(z,\lambda)$ is Owen's function. The properties of this function are:
\begin{enumerate}
\item $-T(z,\lambda)=T(z,-\lambda);$
\item $T(-z,\lambda)=T(z,\lambda);$
\item $2T(z,1)=\Phi(z)\Phi(-z).$
\end{enumerate}
Using the properties of Owen's function, we have immediately  the following ones:
\begin{pr}\label{pr:1}
$1-\Phi(-z;\lambda)=\Phi(z;-\lambda).$
\end{pr}
\begin{pr}\label{pr:2}
$\Phi(z;1)=\Phi(z)^{2}.$
\end{pr}
\begin{pr}\label{pr:3}
$\Phi(z;\lambda)+\Phi(z;-\lambda)=2\Phi(z).$
\end{pr}

The class of skew-normal distributions can be generalized by the inclusion of the location and scale parameters which we identify as $\xi$ and $\psi>0$. Thus if $X\sim SN(\lambda)$ then $Y=\xi+\psi X$ is a skew-normal variable with parameters $\xi,\psi,\lambda$. We denote $Y$ by $Y\sim SN(\xi,\psi,\lambda)$.

\subsection{The Balakrishnan skew-normal density and its generalization}
Balakrishnan (2002) proposed a generalization of standard skew-normal distribution as follows:
\begin{df}
A random variable $X$ has Balakrishnan skew-normal distribution, denoted by $SNB_{n}(\lambda)$, if it has the following  density function, with $n \in \mathbb{N}$,
\begin{equation}
f_{n}(x;\lambda)=c_{n}(\lambda)\phi(x)\Phi(\lambda x)^{n},\quad x\in\mathbb{R},\;\lambda\in\mathbb{R}.\label{eq:bala}
\end{equation}
The coefficient $c_{n}(\lambda)$, which is a function of n and the parameter $\lambda$, is given by $$c_{n}(\lambda)=\frac{1}{\int_{-\infty}^{\infty}\phi(x)\Phi(\lambda x)^{n}dx}=\frac{1}{E\left( \Phi(\lambda U)^{n}\right) },$$ where $U\sim N(0,1)$.
\end{df}
For $n=0$ and $n=1$, the above density reduces to the standard normal and the skew-normal distribution, respectively.\\
For $n=2$ a random variable X with the density \eqref{eq:bala} is denoted by $X\sim NSN(\lambda)$ with $c_{2}(\lambda)=\frac{\pi}{\arctan{\sqrt{1+2\lambda^{2}}} }$ (Sharafi and Behboodian (2006)).

The class of Balakrishnan skew-normal can be generalized by the inclusion of the location and scale parameters which we identify as $\mu$ and $\sigma>0$. Thus if $X\sim SNB_{n}(\lambda)$ then $Y=\mu+\sigma X$ is a Balakrishnan skew-normal variable with parameters $\mu,\sigma,\lambda$. We denote Y by $Y\sim SNB_{n}(\mu,\sigma,\lambda)$.

\begin{rem}\label{rem:1}
Sharafi and Behboodian (2008) have shown that for $\lambda=1$, \eqref{eq:bala} is the density function of the $(n+1)-$th order statistic $X_{(n+1)}$ in a sample of size n+1 from $N(0,1)$. Moreover, for $\lambda=-1$, \eqref{eq:bala} is the density function of the first order statistic $X_{(1)}$ in a sample of size n+1 from $N(0,1)$.
\end{rem}

Recently, Yadegari et al. (2008) introduced the following generalization of the Balakrishnan skew-normal distribution and explained some important properties of this distribution. 
\begin{df}
A random variable $X$ is said to have a generalized Balakrishnan skew-normal distribution, denoted by $GBSN_{n,m}(\lambda)$, with parameters $n,m\in\mathbb{N}$ and $\lambda\in\mathbb{R}$, if its density function has the following form:
\begin{equation}
f_{n,m}(x;\lambda)=\frac{1}{C_{n,m}(\lambda)}\phi(x)\Phi(\lambda x)^{n}\left( 1-\Phi(\lambda x)\right)^{m}\quad x\in\mathbb{R},\label{eq:gbsn}
\end{equation}
where $C_{n,m}(\lambda)=\sum_{i=0}^{m}\binom{m}{i}(-1)^{i}\int_{-\infty}^{\infty}\phi(x)\Phi(\lambda x)^{n+i}dx$.
\end{df}
For $m=0$ this density reduces to the Balakrishnan skew-normal.
\begin{rem}\label{rem:3}
Let $X_{1}, \cdots, X_{n}$ be a random sample from a $N(0,1)$. Then the j-th order statistic is a $GBSN_{j-1,n-j}(1)$, with $j=1,\cdots,n$.
In this case we have that
\begin{equation} C_{j-1,n-j}(1)=\sum_{i=0}^{n-j}\binom{n-j}{i}(-1)^{i}\int_{-\infty}^{\infty}\phi(x)\Phi(x)^{j-1+i}dx=\frac{n!}{(j-1)!(n-j)!}.
\end{equation}
\end{rem}
Bahrami et al. (2009) discussed a two-parameter
generalized skew-normal distribution which includes the skew-normal, the Balakrishnan skew-normal and the generalized Balakrishnan skew-normal as special cases.
\begin{df}\label{def:3}
 A random variable $Z$ has a two-parameter Balakrishnan skew-normal distribution with parameters $\lambda_{1},\lambda_{2}\in\mathbb{R}$, denoted by $T_{n,m}(\lambda_{1},\lambda_{2})$, if its pdf is
\begin{equation}
f_{n,m}(z; \lambda_{1}, \lambda_{2})=\frac{1}{c_{n,m}(\lambda_{1}, \lambda_{2})}\phi(z) \Phi(\lambda_{1}z) ^{n} \Phi(\lambda_{2}z) ^{m},\quad z\in\mathbb{R},
\end{equation}
and $n$, $m$ are non-negative integer numbers.
The coefficient $c_{n,m}(\lambda_{1},\lambda_{2})$, which is a function of the parameters  $n,m,\lambda_{1}$ and $\lambda_{2}$, is given by $$c_{n,m}(\lambda_{1},\lambda_{2}) = E\left[ \Phi(\lambda_{1}X)^{n} \Phi(\lambda_{2}X) ^{m}\right],$$
 where $X\sim N(0, 1)$.
\end{df}
The following properties are direct consequence of definition \eqref{def:3}.\\
\textbf{Properties of $T_{n,m}(\lambda_{1},\lambda_{2})$}:
\begin{enumerate}
\item $TBSN_{1,1}(\lambda_{1}, 0) = SN(\lambda_{1})$ and $TBSN_{1,1}(0,\lambda_{2}) = SN(\lambda_{2})$;
\item $TBSN_{n,m}(\lambda , \lambda) = SNB_{n+m}(\lambda)$;
\item $TBSN_{n,m}(\lambda_{1}, 0) = SNB_{n}(\lambda_{1})$ and $TBSN_{n,m}(0, \lambda_{2}) = SNB_{m}(\lambda_{2})$;
\item $TBSN_{n,m}(\lambda,-\lambda) = GBSN_{n,m}(\lambda)$ and
$TBSN_{n,m}(-\lambda, \lambda) = GBSN_{m,n}(\lambda)$;
\item $TBSN_{n,m}(0, 0) = TBSN_{0,0}(\lambda_{1},\lambda_{2}) = N(0, 1)$.
\end{enumerate}

\begin{rem}\label{rem:4}
Let $Z_{1},\cdots,Z_{n}$ i.i.d $N(0, 1)$ and $Z_{(1)}\leq Z_{(2)}\leq \cdots\leq Z_{(n)}$ be the corresponding order statistics, then $Z_{(i)}\sim TBSN_{i-1,n-i}(1,-1)$.
\end{rem}
The location-scale two-parameter Balakrishnan skew-normal distribution
is defined as the distribution of $Y =\mu +\sigma X$, where $X\sim TBSN_{n,m}(\lambda_{1}, \lambda_{2})$. Hence $\mu\in\mathbb{R}$ and $\sigma>0$, are the location and scale parameters, respectively. We denote $Y$ by $Y\sim TBSN_{n,m}(\mu,\sigma,\lambda_{1}, \lambda_{2})$.

In the rest of the paper we denote by $\phi(.;\lambda)$ the density function of $SN(\lambda)$ and by $f_{n,m}(\cdot;\lambda_{1},\lambda_{2})$ the density function of $TBSN_{n,m}(\lambda_{1},\lambda_{2})$.

\section{Some extensions of the Beta distribution}
In this Section we introduce two families of distributions which generalize the Beta one.
\subsection{The generalized beta of the first type}
We recall the following definition due to McDonald (1984).
\begin{df}
A variable X is said to have a generalized beta distribution of the first kind with positive parameters $a, b, p$ and $q$ if its density is given by
\begin{equation}
g(x;a,b,p,q)=\frac{p x^{ap-1}\left( 1-\left( \frac{x}{q}\right) ^{p}\right)^{b-1}}{q^{ap}B(a,b)}\quad  \textrm{with}\quad  0\leq x\leq q.
\end{equation}
\end{df}
If $p=1$ and $q=1$ the variable $X$ is a Beta of the first kind with parameters $a$ and $b$.\\
For $q=1$ and $a=1$ the variable $X$ is said to have a {\it  Kumaraswamy distribution} with parameters $p$ and $b$ (Kumaraswamy (1980)).
\subsection{The Beta generated distribution}
Starting from the beta distribution, Jones (2004) defined a new family of probability distributions, called Beta generated distribution.  Following the notation of Jones, the class of beta-generated distributions is defined as follows.

\begin{df}
 Let $F(\cdot)$ be a continuous distribution function with density function $f(\cdot)$. The univariate family of distributions generated by $F(\cdot)$, called beta generated distribution, with parameters $a,b>0$, has pdf
\begin{equation}
g_{F}^{B}(x;a,b)=\frac{1}{B(a,b)}(F(x))^{a-1}(1-F(x))^{b-1}f(x)\label{eq:2}
\end{equation}
where $B(a,b)$ is the complete beta function.
\end{df}

Thus, this family of distributions has distribution function given by:
\begin{equation}\label{eq:10}
G_{F}^{B}(x;a,b)=I_{F(x)}(a,b),\quad a,\;b>0
\end{equation}
where the function $I_{F(x)}$ denotes the incomplete beta ratio defined by
\begin{equation}\label{eq:11}
I_{y}(a,b)=\frac{B_{y}(a,b)}{B(a,b)}
\end{equation}
with
\begin{equation}\label{eq:12}
B_{y}(a,b)=\int_{0}^{y}z^{a-1}(1-z)^{b-1}dz,\quad 0<y\leq 1
\end{equation}
 the incomplete beta function.
Replacing \eqref{eq:11} and \eqref{eq:12} in  \eqref{eq:10}, we get that this family of distributions has distribution function
\begin{equation}
G_{F}^{B}(x;a,b)=\frac{1}{B(a,b)}\int_{0}^{F(x)}z^{a-1}(1-z)^{b-1}dz\label{eq:1}.
\end{equation}

\begin{rem}\label{rem:5}
Let $f(\cdot)$ unimodal and continuously differentiable, if $a,b\geq1$ then $g_{F}^{B}(\cdot,a,b)$ is also unimodal. The strong unimodality, i.e. log-concavity, of $f(\cdot)$ implies strong unimodality of $g_{F}^{B}(\cdot{,a,b})$.
\end{rem}

Eugene et al. (2002) studied in details the family of beta-normal distribution (BN) and
discussed its properties. Recently Pescim et al. (2010) proposed the beta half-normal (BHN) distribution to extend the half-normal (HN) distribution. We now recall the definitions of Beta-normal distribution and Beta half-normal distribution.
\subsubsection{Beta-normal distribution}
When in \eqref{eq:2} $F(x)$ is the normal  distribution function with parameters $\mu$ and $\sigma$ we have the \textit{Beta-normal family} with distribution function
\begin{equation}
G_{\Phi(\frac{x-\mu}{\sigma})}^{B}(x;a,b)=\frac{1}{B(a,b)}\int_{0}^{\Phi(\frac{x-\mu}{\sigma})}z^{a-1}(1-z)^{b-1}dz
\end{equation}
and corresponding probability density function
\begin{equation}
g_{\Phi(\frac{x-\mu}{\sigma})}^{B}(x;a,b)=\frac{1}{B(a,b)}\left( \Phi\left( \frac{x-\mu}{\sigma}\right) \right) ^{a-1}\left( 1-\Phi\left( \frac{x-\mu}{\sigma}\right) \right) ^{b-1}\sigma^{-1}\phi\left( \frac{x-\mu}{\sigma}\right),
\end{equation}
where $\sigma^{-1}\phi\left( \frac{x-\mu}{\sigma}\right)$ and $\Phi\left( \frac{x-\mu}{\sigma}\right)$ are the normal density and distribution with parameters $\mu$ and $\sigma$.\\
A random variable $X$ with \textit{Beta-normal distribution} with vector of parameters $\xi =(0,1,a,b)$ is denoted by $X \sim BN(a,b).$

\subsubsection{Beta half-normal distribution}
Let $F(x)=2\Phi(x)-1$, with $x>0$, the distribution function of the half normal distribution. By using $F(x)$ in \eqref{eq:2} the density function of the \textit{Beta half-normal distribution} (BHN) is given by
\begin{equation}
g_{2\Phi(x)-1}^{B}(x;a,b)=\frac{2^{b}}{B(a,b)}( 2\Phi(x)-1) ^{a-1}(1-\Phi(x) )^{b-1}\phi( x),\quad x>0
\end{equation}
and the relative distribution function is 
\begin{equation}
G_{2\Phi(x)-1}^{B}(x;a,b)=\int_{0}^{2\Phi(x)-1}\frac{2^{b}}{B(a,b)}( 2\Phi(t)-1) ^{a-1}(1-\Phi(t) )^{b-1}\phi( t)dt,\quad x>0.
\end{equation}
When $X$ is a random variable following the BHN distribution, it is denoted by $X\sim BHN(a,b)$.

\section{A new generalization of the skew-normal distribution: the Beta skew-normal}
In this Section we define the Beta skew-normal class and we present some of its properties.
Replacing in \eqref{eq:2} $F(x)$ by $\Phi(x;\lambda)$, we obtain the \textit{Beta skew-normal distribution}, with distribution function given by
\begin{equation}
G_{\Phi(x;\lambda)}^{B}(x;\lambda,a,b)=\frac{1}{B(a,b)}\int_{0}^{\Phi(x;\lambda)}z^{a-1}(1-z)^{b-1}dz
\end{equation}
and probability density function
\begin{equation}
g_{\Phi(x;\lambda)}^{B}(x;\lambda,a,b)=\frac{2}{B(a,b)}(\Phi(x;\lambda))^{a-1}(1-\Phi(x;\lambda))^{b-1}\phi(x)\Phi(\lambda x).\label{eq:a}
\end{equation}
Throughout this paper, we denote the Beta skew-normal distribution with  vector of parameters $\xi =(\lambda,a,b)$ by $BSN(\lambda,a,b).$ The class of the Beta skew-normal can be generalized by the inclusion of the location and scale parameters which we identify as $\mu$ and $\sigma>0$. Thus if $X\sim BSN(\lambda,a,b)$ then $Y=\mu+\sigma X$ is a Beta skew-normal with vector of parameters $\xi=(\mu,\sigma,\lambda,a,b)$. We denote Y by $Y\sim BSN(\mu,\sigma,\lambda,a,b)$.

We now present some properties concerning the $BSN(\lambda,a,b).$\\
\textbf{Properties of $BSN(\lambda,a,b)$}:
\begin{itemize}
\item[a.]$g_{\Phi(x;\lambda)}^{B}(x;\lambda,1,1)=\phi(x;\lambda)$ for all $x\in\mathbb{R}$, i.e. $BSN(\lambda,1,1)=SN(\lambda)$.
\item[b.]$g_{\Phi(x;0)}^{B}(x;0,a,b)=g_{\Phi(x)}^{B}(x;a,b)$ for all $x\in\mathbb{R}$, i.e. $BSN(0,a,b)=BN(a,b)$.
\item[c.] $g_{\Phi(x;0)}^{B}(x;0,1,1)=\phi(x)$ for all $x\in\mathbb{R}$, i.e. $BSN(0,1,1)=N(0,1)$.
\item[d.] $g_{\Phi(x;1)}^{B}(x;1,\frac{1}{2},1)=\phi(x)$ for all $x\in\mathbb{R}$, i.e. $BSN(1,\frac{1}{2},1)=N(0,1)$.
\item[e.] $g_{\Phi(x;-1)}^{B}(x;-1,1,\frac{1}{2})=\phi(x)$ for all $x\in\mathbb{R}$, i.e. $BSN(-1,1,\frac{1}{2})=N(0,1)$.
\item[f.] If $X\sim BSN(\lambda,a,b)$ then $-X\sim BSN(-\lambda,b,a)$.
\item[g.] If $X\sim BSN(\lambda,a,b)$ then $Y=\Phi(X;\lambda)$ is a $Beta(a,b)$.
\item[h.] If $X\sim BSN(\lambda,a,b)$ then $Y=1-\Phi(X;\lambda)$ is a $Beta(b,a)$.
\item[i.] As $\lambda\rightarrow +\infty$, $g_{\Phi(x;\lambda)}^{B}(x;\lambda,a,b)$ tends to the Beta half-normal density.
\end{itemize}
\begin{rem}
Properties from a to e establish that the family of $BSN(\lambda,a,b)$ contains the standard normal distribution, the skew-normal distribution and the Beta-normal distribution as special cases.
\end{rem}
\begin{proof}
The proof follows directly from \eqref{eq:a} and from elementary properties of the skew-normal distribution.
 \end{proof}
The BSN distribution is easily simulated using Property~g as follows: if $Y$ has a beta distribution with parameters $a$ and $b$, then the variable $X=\Phi^{-1}(Y;\lambda)$ has $BSN(\lambda,a,b)$ distribution, where $\Phi^{-1}(.;\lambda)$ is the quantile function of the skew-normal distribution.
In Figure~1 are plotted random samples generated by the BSN distribution for some $a$, $b$, $\lambda$ with the respective curve of the density function obtained using R.\\
By Property~f we can deduce the following proposition.
\begin{prop}
Let $X\sim BSN(\lambda,a,b)$ and $Y\sim BSN(-\lambda,b,a)$. We have the following statements:
\begin{enumerate}
\item $E_{X}(X)=-E_{Y}(Y)$;
\item $var_{X}(X)=var_{Y}(Y)$;
\item $\gamma_{1}(X)=-\gamma_{1}(Y)$;
\item $\gamma_{2}(X)=\gamma_{2}(Y)$;
\end{enumerate}
with $\gamma_{1}$ and $\gamma_{2}$ we indicate the skewness and the kurtosis, respectively.
\end{prop}
Now we find the moment generating function of $X$ which has density \eqref{eq:a}.
\begin{pr}
The moment generating function of $X\sim BSN(\lambda,a,b)$ is given by
\begin{equation}
M_{X}(t)=\frac{2}{B(a,b)}e^{\frac{t^2}{2}}E_{Z}\left[ (\Phi(Z;\lambda))^{a-1}(1-\Phi(Z;\lambda))^{b-1}\Phi(\lambda Z)\right] 
\end{equation}
where $Z\sim N(t,1)$.
\end{pr}

We have the following recursion formula:
\begin{pr}
Let $k\in \mathbb{N}$ and $k\geq 2$. If $X\sim BSN(\lambda,a,b)$, with $a,b>1$ then 
\begin{align*}
E_{X}(X^{k})&=(k-1)E_{X}(X^{k-2})+\lambda E_{X}\left[  X^{k-1}\frac{\phi(\lambda X)}{\Phi(\lambda X)}\right]  +\\
&+(a+b-1)E_{U}\left[  U^{k-1}\phi(U;\lambda)\right]  
-(a+b-1)E_{V}\left[  V^{k-1}\phi(V;\lambda)\right]  
\end{align*}
where $U\sim BSN(\lambda,a-1,b)$ and $V\sim BSN(\lambda,a,b-1)$ are independent random variables.
\end{pr}
\begin{proof}
The proof follows easily from application of the formula for integration by parts and by using the well note $\frac{\partial\phi(x)}{\partial x}=-x\phi(x)$ (see Arnold et al.(1992)).
\end{proof}

The Beta skew-normal density is in general asymmetric (see Figures~2 and ~3). We have a partial result concerning symmetry:

\begin{prop}
If $a=b$ and $BSN(\lambda,a,b)$ is symmetric about 0 then $\lambda=0$.
\end{prop}
\begin{proof}
We consider the density of a random variable $X\sim BSN(\lambda,a,a)$:
\begin{align*}
g_{\Phi(x;\lambda)}^{B}(-x;\lambda,a,a)&=
\frac{2}{B(a,a)}\phi(x)\Phi(-\lambda x)(1-\Phi(x;-\lambda))^{a-1}(\Phi(x;-\lambda))^{a-1}&
\end{align*}
this is equal to $g_{\Phi(x;\lambda)}^{B}(x;\lambda,a,a)$ if $\Phi(\lambda x)=\Phi(-\lambda x)$ and $\Phi(x;\lambda)=\Phi(x;-\lambda)$. However for Property~\ref{pr:3} we find that $\Phi(x;\lambda)=\Phi(x)$ which implies that $\lambda=0$.
\end{proof}

\begin{rem}
Eugene et al. (2002) have shown that the $BN(a,b)=BSN(0,a,b)$ is symmetric about 0 when $a=b$.
\end{rem}

\begin{rem}
From Remark~\ref{rem:5} we know that, if $a,b\geq 1$, the density \eqref{eq:a} is strongly unimodal, i.e $\log{g^{B}_{\Phi(x;\lambda)}(x;\lambda,a,b)}$ is a concave function of $x$. We don't have general results for  $a$ and/or $b$ $<1$. A numerical study has shown that, when at least one of the two  parameters $a$ and $b$ is closed to zero $(0.10, 0.20)$, the density can be bimodal (see figure~3). Numerically Eugene et al. (2002) observed that the BN is bimodal when both parameters $a$ and $b$ are less then $0.214$.
\end{rem}

Moments of the BSN cannot be evaluated exactly. We have computed them numerically using the software R. In Table ($1$) we have reported the values of the mean $\mu_{BSN}$, standard deviation $\sigma_{BSN}$, skewness $\gamma_{1}$ and kurtosis $\gamma_{2}$ for different values of the parameters $a$, $b$ and $\lambda$. From these numerical study we have noted that:
\begin{itemize}
\item for fixed values of $a$ and $b$ the mean $\mu_{BSN}$ and skewness $\gamma_{1}$ are increasing function of $\lambda$;
\item for fixed values of  $b$ and $\lambda$ the mean $\mu_{BSN}$ and  skewness $\gamma_{1}$ are increasing function of $a$;
\item for fixed values of $a$ and $\lambda$ the mean $\mu_{BSN}$ is a decreasing function of $b$.
\end{itemize}
We now give some results concerning the distribution of order statistics from a skew-normal distribution:
\begin{prop}\label{prop:3}
Let $X_{1},\cdots, X_{n}$ be a random sample from a $SN(\lambda)$. Then the j-th order statistic is a $BSN(\lambda,j,n-j+1)$, where $j=1\cdots,n$.
\end{prop}
\begin{proof}
The proof readily follows using the  standard formula of the density of $X_{(i)}$, the $i$-th order statistic of a random sample of size $n$ from the distribution $SN(\lambda)$.
\end{proof}
From Proposition~\ref{prop:3} follows immediately that the family of BSN contains the distributions of the order statistics of the skew-normal distribution.\\
In particular we have the following corollaries:
\begin{cor}\label{cor:1}
Let $X_{1},\cdots, X_{n}$ be a random sample from a $SN(1)$. Then $$X_{(n)}=\max\left\lbrace X_{1},\cdots,X_{n}\right\rbrace$$ is a $BSN(1,n,1)$.
\end{cor}
\begin{cor}\label{cor:2}
Let $X_{1},\cdots, X_{n}$ be a random sample from a $SN(-1)$. Then $$X_{(1)}=\min\left\lbrace X_{1},\cdots,X_{n}\right\rbrace$$ is a $BSN(-1,1,n)$.
\end{cor}
\begin{cor}
Let $X_{(1)}<X_{(2)}<\cdots<X_{(n)}$ be the order statistic from a sample of size $n$ from a $SN(\lambda)$ distribution. Then $X_{(i)}$, $i=1,\cdots,n$, has log-concave density.
\end{cor}
\begin{proof}
From Property~d of Section~2 we know that $X_{i}$ has a log-concave density. We conclude the proof using the following result due to Gupta (2004): \textit{Suppose $X_{(1)}<X_{(2)}<\cdots<X_{(n)}$ be the order statistic from a sample of size n from a distribution having a log-concave density function. Then $X_{(i)}$, $i=1,\cdots,n$, has log-concave density.}
\end{proof}
We now derive other properties of the BSN distribution.
\begin{teo}\label{th:1}
Let $X\sim BSN(\lambda,a,b)$ be independent of a random sample $\left(Y_{1},\cdots,Y_{n}  \right)$ from $SN(\lambda)$, then $X|\left( Y_{(n)}\leq X\right) \sim BSN(\lambda,a+n,b)$ and $X|\left( Y_{(1)}\geq X\right) \sim BSN(\lambda,a,b+n)$, where $Y_{(n)}$ and $Y_{(1)}$ are the largest and the smallest order statistics, respectively.
\end{teo}
\begin{proof}
If $W=X|\left( Y_{(n)}\leq X\right) $, then we have
\begin{equation}\label{eq:c}
P(W\leq w)=\frac{\int_{-\infty}^{w}\left( \Phi(x;\lambda)\right)^{n}\frac{2}{B(a,b)}\phi(x)\Phi(\lambda x)\left( \Phi(x;\lambda)\right)^{(a-1)}\left( 1-\Phi(x;\lambda)\right)^{(b-1)} dx}{P(Y_{(n)}\leq X)}.
\end{equation}
Also
\begin{align*}
P(Y_{(n)}\leq X)&= P(Y_{1}\leq X,\cdots,Y_{n}\leq X)=\\
&=\int_{-\infty}^{\infty}\left( \Phi(x;\lambda)\right)^{n}\frac{2}{B(a,b)}\phi(x)\Phi(\lambda x)\left( \Phi(x;\lambda)\right)^{(a-1)}\left( 1-\Phi(x;\lambda)\right)^{(b-1)} dx=\\
&=\frac{B(a+n,b)}{B(a,b)}.
\end{align*}
Taking derivative from \eqref{eq:c} with respect to $w$, we obtain the $BSN(\lambda,a+n,b)$ density function, the proof of $X|\left( Y_{(1)}\geq X\right) \sim BSN(\lambda,a,b+n)$ is similar.
\end{proof}
\begin{teo}
If $X\sim BSN(\lambda,a,b)$ is independent of the random sample $\left( U_{1},\cdots,U_{n-1},V_{1},\cdots,V_{m-1}\right)$ from $SN(\lambda)$
then $X|\left( U_{(n-1)}\leq X,V_{(1)}\geq X\right) \sim BSN(\lambda,a+n-1,b+m-1)$, where $U_{(n-1)}$ and $V_{(1)}$ are the largest and the smallest order statistics, respectively.
\end{teo}
\begin{proof}
The proof is quite similar to the one of Theorem \ref{th:1}.
\end{proof}
Theorem~\ref{th:1} can be used to generate $X\sim BSN(\lambda,n,1)$ by extending the acceptance-rejection technique, due to Azzalini, as follows (see Azzalini (1985) and Sharafi and Behboodian (2008)): first we generate a random sample $T,U_{1},U_{2},\cdots,U_{n-1}$ from $SN(\lambda)$, if $\max(U_{1},U_{2},\cdots,U_{n-1})\leq T$ we put $X=T$. Otherwise, we generate a new random sample, until the above inequality is satisfied.

\subsection{Further results concerning the Balakrishnan skew-normal and the Beta skew-normal}
In this Section we present some results concerning the SNB distribution and link the distributions introduced in Section 2.2 with the Beta skew-normal.
First we consider two results about the Balakrishnan skew-normal. We study the distribution of the largest order statistic from $SNB_{m}(1)$ and subsequently the distribution of the smallest order statistic from $SNB_{m}(-1)$. We found that these distributions belong to the family of SNB.
\begin{prop}\label{prop:4}
Let $X_{1},\cdots, X_{n}$ be a random sample from a $SNB_{m}(1)$. Then  $$X_{(n)}=\max\left\lbrace X_{1},\cdots,X_{n}\right\rbrace$$ is a $SNB_{k}(1)$, where $k=n(m+1)-1$.
\end{prop}
\begin{proof}
The proof follows easily using the standard formula for the density of $X_{(n)}$, the largest order statistic of a random sample of size $n$ from the distribution $SNB_{m}(1)$. 
\end{proof}
In particular the following corollaries hold:
\begin{cor}\label{cor:4}
Let $X_{1},\cdots, X_{n}$ be a random sample from a $SN(1)$. Then $$X_{(n)}=\max\left\lbrace X_{1},\cdots,X_{n}\right\rbrace$$  is a $SNB_{2n-1}(1)$.
\end{cor}
\begin{proof}
The skew-normal distribution with parameter $\lambda=1$ is a Balakrishnan skew-normal with parameters $\lambda=1$ and $m=1$.\\
The same result can be established making use of the well-known result for the density of the largest order statistic from the distribution $SN(1)$ and Property~\ref{pr:2}.
\end{proof}
\begin{cor}\label{cor:5}
Let $X_{1},\cdots, X_{n}$ be a random sample from a $SNB_{m}(-1)$. Then  $$X_{(1)}=\min\left\lbrace X_{1},\cdots,X_{n}\right\rbrace$$ is a $SNB_{k}(-1)$, where $k=n(m+1)-1$.
\end{cor}
It follows immediately from corollaries \eqref{cor:1}, \eqref{cor:2}, \eqref{cor:4} and \eqref{cor:5} that the BSN distribution is related to the previous skew-normal generalizations. In fact, its density simplifies to the Balakrishnan skew-normal when $b=1$, $\lambda=1$ and $a\geq 1$ integer (or $a=1$, $\lambda=-1$ and $b\geq 1$ integer). Further, if
 $\lambda=0$ it reduces to the generalized Balakrishnan skew-normal when $a$ and $b$ are both integers. These consideration have been summarized in the following proposition.
\begin{prop}
The BSN distribution satisfies the following properties:
\begin{itemize}
\item $g_{\Phi(x;1)}^{B}(x;1,n,1)=f_{2n-1,m}(x;1,0)$ for all $x\in\mathbb{R}$, i.e. $BSN(1,n,1)=TBSN_{2n-1,m}(1,0)$;
\item $g_{\Phi(x;-1)}^{B}(x;-1,1,m)=f_{n,2m-1}(x;0,-1)$ for all $x\in\mathbb{R}$, i.e. $BSN(-1,1,m)=TBSN_{n,2m-1}(0,-1)$;
\item $g_{\Phi(x;0)}^{B}(x;0,n,m)=f_{n-1,m-1}(x;1,-1)$ for all $x\in\mathbb{R}$, i.e. $BSN(0,n,m)=TBSN_{n-1,m-1}(1,-1)$;
\end{itemize}
where $n,m$ are positive integer numbers.
\end{prop}
Given a random variable $X\sim BSN(\lambda,a,b)$ we are interested in constructing a random variable $Y$ with Kumaraswamy distribution. This goal can be achieved using the below properties which follow easily from Property~g and Property~h of Section~4, respectively.
\begin{pr}
If $X\sim BSN(\lambda,1,b)$ then $Y=\left( \Phi(X;\lambda)\right)^{\frac{1}{a}}$ is a $Kumaraswamy(a,b).$ In particular, if $X\sim SNB_{2b-1}(-1)$ then $Y=\left(1-\Phi(-X)^{2}\right)^{\frac{1}{a}}$ is a $Kumaraswamy(a,b)$.
\end{pr}

\begin{pr}
If $X\sim BSN(\lambda,a,1)$ then $Y=\left(1-\Phi(X;\lambda)\right)^{\frac{1}{b}}$ is a $Kumaraswamy(b,a).$ In particular, if $X\sim SNB_{2a-1}(1)$ then $Y=\left(1-\Phi(X)^{2}\right)^{\frac{1}{b}}$ is a $Kumaraswamy(b,a)$.
\end{pr}
 Recently, Ferreira and Steel (2006) have presented a general approach which allows to generate skew distributions. They show that every univariate continuous skew distribution can be obtained from a ``perturbation'' of a symmetric one as it explained in the following definition:
 \begin{df} A distribution $S$ is said to be a skewed version
 of the symmetric distribution $F$, generated by the skewing
 mechanism $P$, if its pdf is of the form
 \begin{equation}\label{skewed}
 s(y|f,p) = f(y)p[F(y)],\qquad y\in\mathbb{R}
 \end{equation}
 where $f$ and $F$ are the pdf and cdf of a symmetric distribution on the real line, respectively, and $p$ is the pdf of a distribution on $(0,1)$.
 \end{df}
Note that, if $F$ is the standard normal distribution and $p$ on
 $(0, 1)$ is given by
 \begin{equation}
 p(u;\lambda,a,b)=\frac{2}{B(a,b)}\Phi(\lambda \Phi^{-1}(u))\left( \Phi(\Phi^{-1}(u);\lambda)\right)^{a-1} \left( 1-\Phi(\Phi^{-1}(u);\lambda)\right)^{b-1},
 \end{equation}
formula \eqref{skewed} reduces to Beta skew-normal with parameters $\lambda,a,b$. Then the pdf of a Beta skew-normal with parameters $\lambda,a,b$ can be seen as a weighted version of $\phi(y)$, with weight function given by $p\left( \Phi(y);\lambda,a,b\right) $.
\newpage
\noindent BIBLIOGRAPHY
\vskip 3mm

\noindent  Arellano-Valle, R. B.,  and Azzalini, A. (2006). On the unification of families of skew-normal distributions. {\it Scand. J. Statist.}, {\bf 33}, 561--574.
\vskip 3mm
\noindent Arnold, B. C., Balakrishnan, N.,  Nagaraya, H. N. (1992). {\it A first course in Order statistics}. New York: Wiley.
\vskip 3mm
\noindent Arnold, B. C., and Beaver, R. J. (2002). Skewed multivariate models related to hidden truncation and/or selective reporting (with discussion). {\it Test}, {\bf 11(1)}, 7--54.
\vskip 3mm
\noindent  Azzalini, A. (1985). A Class of Distributions Which Includes the Normal Ones. {\it Scandinavian Journal of Statistics}, {\bf 12(2)}, 171--178.
\vskip 3mm
\noindent Bahrami, W., Agahi, H., Rangin, H. (2009).
A Two-parameter Balakrishnan
Skew-normal Distribution. {\it J. Statist. Res. Iran}, {\bf 6}, 231--242.
\vskip 3mm
\noindent Balakrishnan, N. (2002). Discussion of ``Skewed multivariate models related to hidden truncation and/or selective reporting''. {\it Test}, {\bf 11}, 37--39.
\vskip 3mm
\noindent  Cooray, K. and Ananda, M. M. A. (2008). A generalization of the half-normal distribution whit applications to lifetime data. {\it Communication in Statistics} \& {\it Theory and Methods}, {\bf 37}, 1323--1337.
\vskip 3mm
\noindent  David, H. A., Nagaraya, H. N. (2003). {\it Order statistics}. Hoboken, NJ: Wiley.
\vskip 3mm
\noindent  Eugene, N., Lee, C., Famoye, F. (2002). Beta-normal distribution and its applications. {\it Communications in Statistics - Theory methods}, {\bf 31(4)}, 497--512.
\vskip 3mm
\noindent Ferreira, J. T. A. S, Steel, M. F. J. (2006). A constructive representation of univariate skewed distributions, {\it Journal of the American Statistical Association}, {\bf 101 (474)} , 823--829.
\vskip 3mm
\noindent  Gupta, R. C., Gupta, R. D. (2004). Generalized skew-normal model. {\it Test} , {\bf 13(2)}, 501--524.
\vskip 3mm
\noindent  Jamalizadeh, A., Balakrishnan, N. (2009). Order statistics from trivariate normal and $t_{\nu}-$distributions in terms of generalized skew-normal and skew-$t_{\nu}$ distributions. {\it Journal of Statistical Planning and Inference}, {\bf 139}, 3799--3819.
\vskip 3mm
\noindent  Jones, M. C. (2004). Families of distributions of order statistics. {\it Test}, {\bf 13(1)}, 1--43.
\vskip 3mm
\noindent Kumaraswamy, P. (1980). A generalized probability density function for double-bounded random processes. {\it Journal of Hydrology}, {\bf 46}, 79--88.
\vskip 3mm
\noindent  Loperfido, N. (2006). Statistical implications of selectively reported inferential results. {\it Statistics} \& {\it Probability Letters}, {\bf 56}, 13--22.
\vskip 3mm
\noindent  Loperfido, N. (2008). A note on skew-elliptical distributions and linear functions of order statistics. {\it Statistics} \& {\it Probability Letters}, {\bf 78}, 3184--3186.
\vskip 3mm
\noindent McDonald, J. B. (1984). Some generalized functions for the size distribution of income. {\it Econometrica}, {\bf 52}, 647--664.
\vskip 3mm
\noindent  Pescim, R. R. Demétrio, C. G. B., Cordeiro G. M. E., Ortega, M. M.,Urbanoa, M. R., (2010). The beta generalized half-normal distribution. {\it Computational Statistics and Data Analysis}, {\bf 54(4)}, 945--957.
\vskip 3mm
\noindent  Sharafi, M., Behboodian, J. (2006). A new skew-normal density. {\it J. Statist. Res. Iran}, {\bf 3}, 47--61.
\vskip 3mm
\noindent  Sharafi, M., Behboodian, J. (2008). The Balakrishnan skew-normal density. {\it Statistical Papers}, {\bf 49}, 769--778.
\vskip 3mm
\noindent  Yadegari, I., Gerami, A., Khaledi, M.J. (2008). A generalization of the Balakrishnan skew-normal distribution. {\it Statistics and Probabilty Letters}, {\bf 78}, 1165--1167.


\begin{table}[htbp]
\centering
\resizebox{1\textwidth}{!}{
\subtable[\label{tab:1}]{
\begin{tabular}{|ccccccc|}
\hline
\multicolumn{1}{|c}{$a$} & \multicolumn{1}{c}{$b$} & \multicolumn{1}{c}{$\lambda$} & \multicolumn{1}{c}{$\mu_{BSN}$} & \multicolumn{1}{c}{$\sigma_{BSN}$} & \multicolumn{1}{c}{$\gamma_{1}$} & \multicolumn{1}{c|}{$\gamma_{2}$}\\
\hline
$0.25$ & $0.25$ & $-10$ & $-1.1579$ & $1.4029$ & $ -1.1329$ & $ 3.7648$\\
$$ & $$ & $-1$ & $ -0.6501$ & $ 1.9679$ & $ -0.2378$ & $2.7777$\\
$$ & $$ & $0$ & $0$ & $ 2.3382$ & $ -0.0004$ & $ 2.6217$\\
$$ & $$ & $1$ & $0.6484$ & $1.9649$ & $ 0.2306$ & $ 2.1362$\\
$$ & $$ & $10$ & $ 1.1580$ & $ 1.4027$ & $1.1329$ & $3.7632$\\
\hline
$0.25$ & $0.5$ & $-10$ & $-1.5906$ & $ 1.3469
$ & $ -0.7185$ & $ 2.7580$\\
$$ & $$ & $-1$ & $ -1.4424$ & $1.6716$ & $-0.3284$ & $ 3.0202$\\
$$ & $$ & $0$ & $ -0.9631$ & $ 1.9061$ & $-0.0849$ & $2.8029$\\
$$ & $$ & $1$ & $ -0.1772$ & $1.5265$ & $0.0938$ & $2.8543$\\
$$ & $$ & $10$ & $ 0.5446$ & $0.8728$ & $ 1.5054$ & $5.1988$\\
\hline
$0.5$ & $0.25$ & $-10$ & $-0.5447$ & $0.8727$ & $-1.5061$ & $5.2003$\\
$$ & $$ & $-1$ & $0.1773$ & $1.5265$ & $-0.0938$ & $2.8541$\\
$$ & $$ & $0$ & $0.9625$ & $1.9051$ & $0.0819$ & $2.7927$\\
$$ & $$ & $1$ & $1.4411$ & $1.6694$ & $0.3203$ & $2.9849$\\
$$ & $$ & $10$ & $1.6339$ & $1.3974$ & $0.8434$ & $3.2655$\\
\hline
$0.5$ & $0.5$ & $-10$ & $-0.8979$ & $0.8874$ & $-0.9703$ & $3.3176$\\
$$ & $$ & $-1$ & $-0.5882$ & $1.2659$ & $-0.1811$ & $2.9514$\\
$$ & $$ & $0$ & $0$ & $1.5253$ & $0$ & $2.8615$\\
$$ & $$ & $1$ & $0.5882$ & $1.2659$ & $0.1811$ & $2.9514$\\
$$ & $$ & $10$ & $0.9179$ & $0.9153$ & $1.0703$ & $3.7747$\\

\hline
$0.5$ & $1$ & $-10$ & $-1.3018$ & $0.9148$ & $-0.8262$ & $3.4815$\\
$$ & $$ & $-1$ & $-1.1664$ & $1.0704$ & $-0.3085$ & $3.1159$\\
$$ & $$ & $0$ & $-0.7043$ & $1.2479$ & $-0.1372$ & $2.9831$\\
$$ & $$ & $1$ & $0$ & $0.9999$ & $0$ & $2.9999$\\
$$ & $$ & $10$ & $0.4873$ & $0.5778$ & $1.3199$ & $4.8561$\\
\hline
\end{tabular}
}\qquad\qquad
\subtable[\label{tab:2}]{
\begin{tabular}{|ccccccc|}
\hline
\multicolumn{1}{|c}{$a$} & \multicolumn{1}{c}{$b$} & \multicolumn{1}{c}{$\lambda$} & \multicolumn{1}{c}{$\mu_{BSN}$} & \multicolumn{1}{c}{$\sigma_{BSN}$} & \multicolumn{1}{c}{$\gamma_{1}$} & \multicolumn{1}{c|}{$\gamma_{2}$}\\
\hline
$0.5$ & $10$ & $-10$ & $-2.3678$ & $0.7314$ & $-0.7505$ & $3.7967$\\
$$ & $$ & $-1$ & $-2.3617$ & $0.7389$ & $-0.7188$ & $3.7849$\\
$$ & $$ & $0$ & $-2.0809$ & $0.8033$ & $-0.6173$ & $3.5736$\\
$$ & $$ & $1$ & $-1.0893$ & $0.6117$ & $-0.5642$ & $3.4799$\\
$$ & $$ & $10$ & $-0.0182$ & $0.1429$ & $0.3706$ & $3.8635$\\
\hline
$1$ & $0.5$ & $-10$ & $-0.4873$ & $0.5777$ & $-1.3200$ & $4.8570$\\
$$ & $$ & $-1$ & $0$ & $1$ & $0$ & $3$\\
$$ & $$ & $0$ & $0.7043$ & $1.2479$ & $0.1372$ & $2.9831$\\
$$ & $$ & $1$ & $1.1664$ & $1.0704$ & $0.3086$ & $3.1161$\\
$$ & $$ & $10$ & $1.3018$ & $0.9148$ & $0.8262$ & $3.4814$\\
\hline
$1$ &  $1$ & $-10$ & $ -0.7939$ & $ 0.6080$ & $ -0.9556$  & $ 3.8232$\\
$$ & $$ & $-1$ & $ -0.5642$ & $0.8256$ & $-0.1369$ & $3.0617$\\
$$ & $$ & $0$ & $0$ & $1$ & $0$ & $3$\\
$$ & $$ & $1$ & $ 0.5642$ & $0.8256$ & $ 0.1369$ & $ 3.0617$\\
$$ & $$ & $10$ & $ 0.7939$ & $ 0.6080$ & $ 0.9556$ & $3.8232$\\
\hline
$10$ & $1$ & $-10$ & $-0.0839$ & $0.1364$ & $-0.7082$ & $ 4.2018$ \\
$$ & $$ & $-1$ & $0.6744$ & $0.4536$ & $0.3597$ & $3.2722$\\
$$ & $$ & $0$ & $1.5388$ & $0.5868$ & $0.4099$ & $3.3314$\\
$$ & $$ & $1$ & $1.8675$ & $ 0.5251$ & $ 0.5005$ & $ 3.4685$\\
$$ & $$ & $10$ & $1.8807$ & $0.5124$ & $0.5744$ & $ 3.5243$\\
\hline
$1$ & $10$ & $-10$ & $-1.8807$ & $0.5124$ & $-0.5744$ & $ 3.5243$\\
$$ & $$ & $-1$ & $-1.8675$ & $0.5251$ & $-0.5005$ & $3.4685$\\
$$ & $$ & $0$ & $ -1.5388$ & $0.5868$ & $ -0.4099$ & $3.3314$\\
$$ & $$ & $1$ & $-0.6744$ & $0.4536$ & $-0.3597$ & $3.2722$\\
$$ & $$ & $10$ & $0.0839$ & $0.1364
$ & $0.7082$ & $4.2018$\\
\hline
\end{tabular}}}
\caption{The first moment, the standard deviation, the skewness and the kurtosis of $BSN(\lambda,a,b)$ for different values of $a$, $b$ and $\lambda$ }
\end{table}

\begin{figure}[tb]\label{fig:1}
\centering
\subfigure{\includegraphics[width=0.49\textwidth]{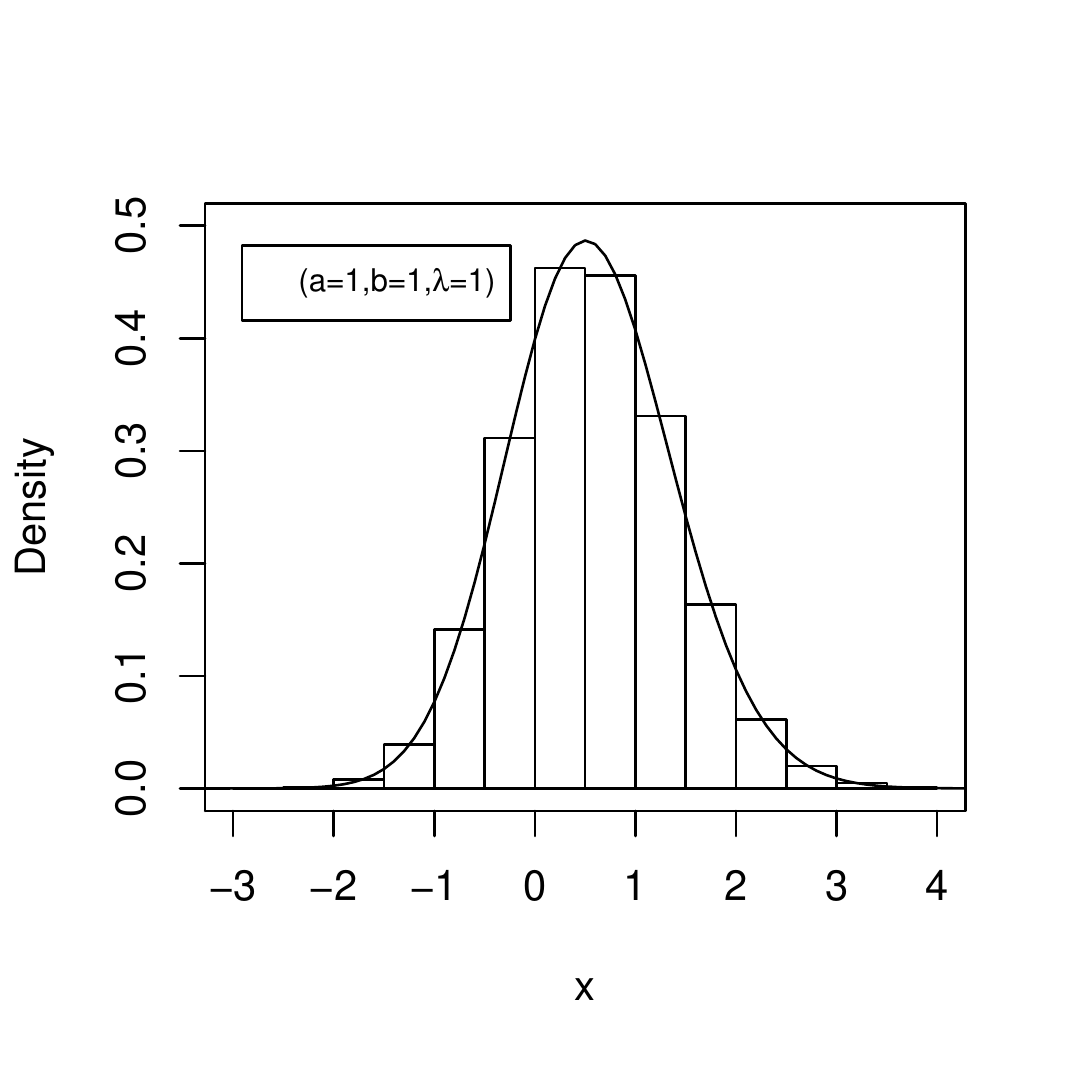}}
\subfigure{\includegraphics[width=0.49\textwidth]{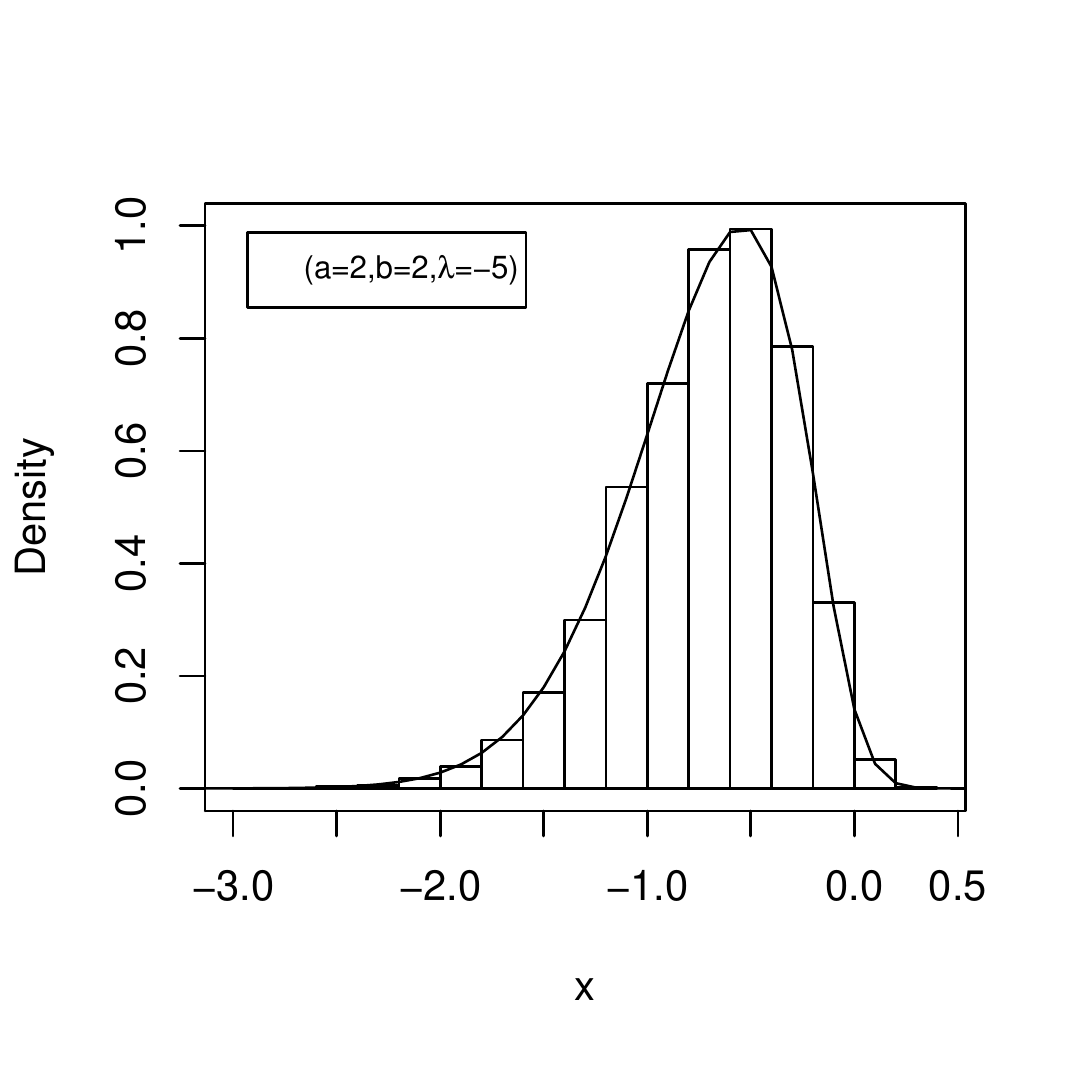}}
\subfigure{\includegraphics[width=0.6\textwidth]{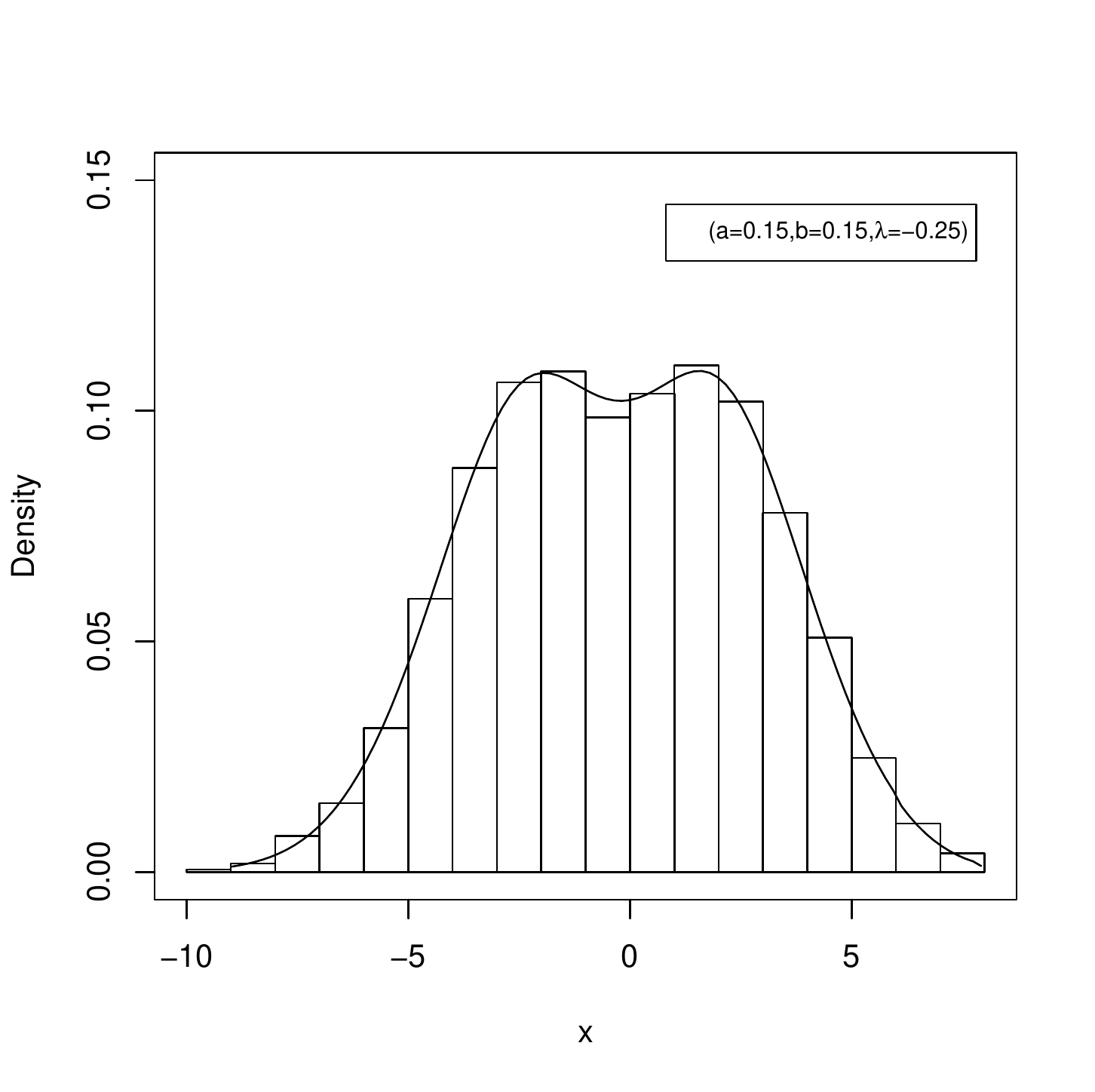}}
\caption{Generated samples of the BSN distribution for some values of $\lambda,a$ and $b$}
\end{figure}

\begin{figure}[!p]\label{fig:2}
\centering
\includegraphics[width=0.5\textwidth]{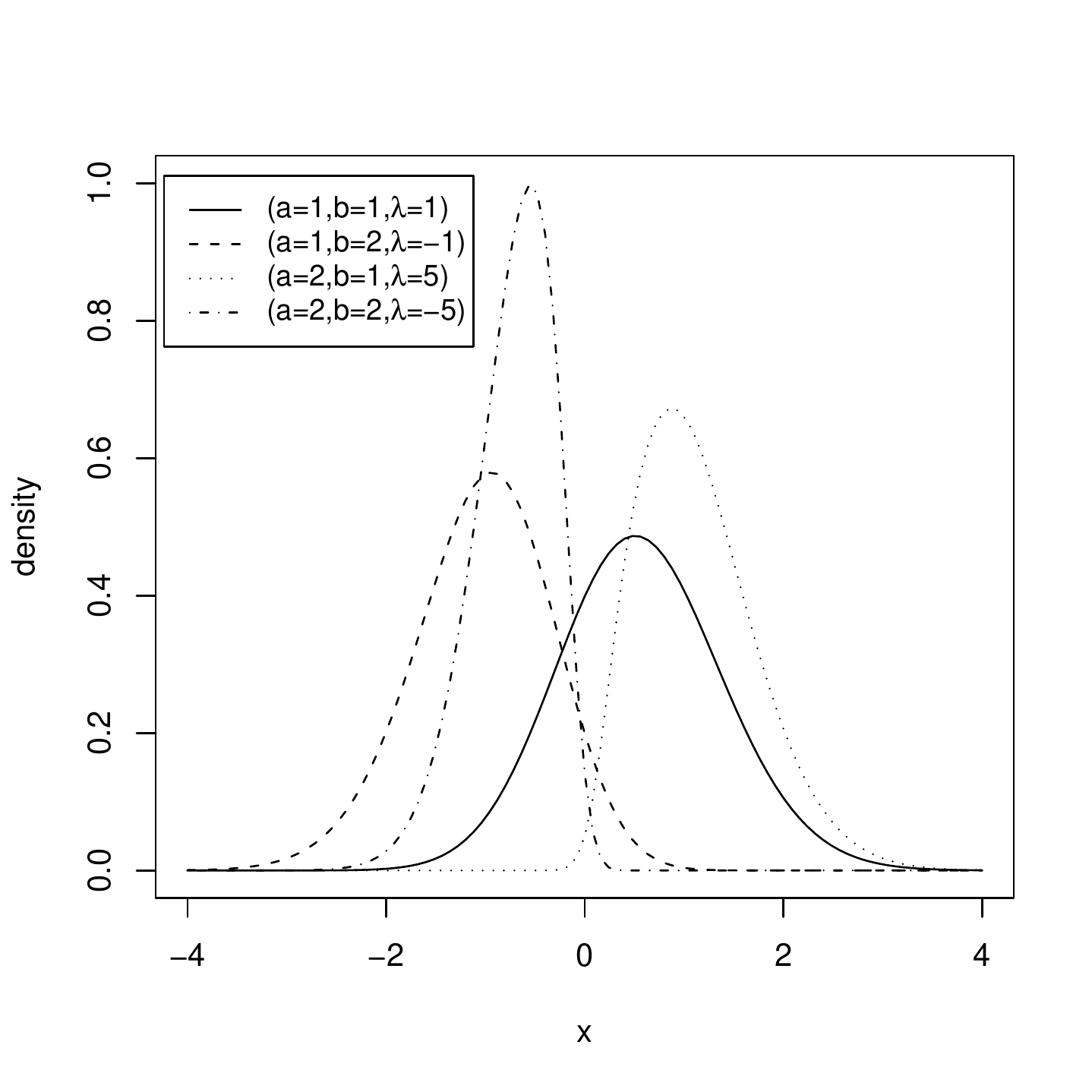}  
\caption{\textit{The Beta skew-normal $BSN(\lambda,a,b)$ for values of $a,b\geq 1$}}
\end{figure}

\begin{figure}[!p] \label{fig:3}
\centering
\includegraphics[width=0.5\textwidth]{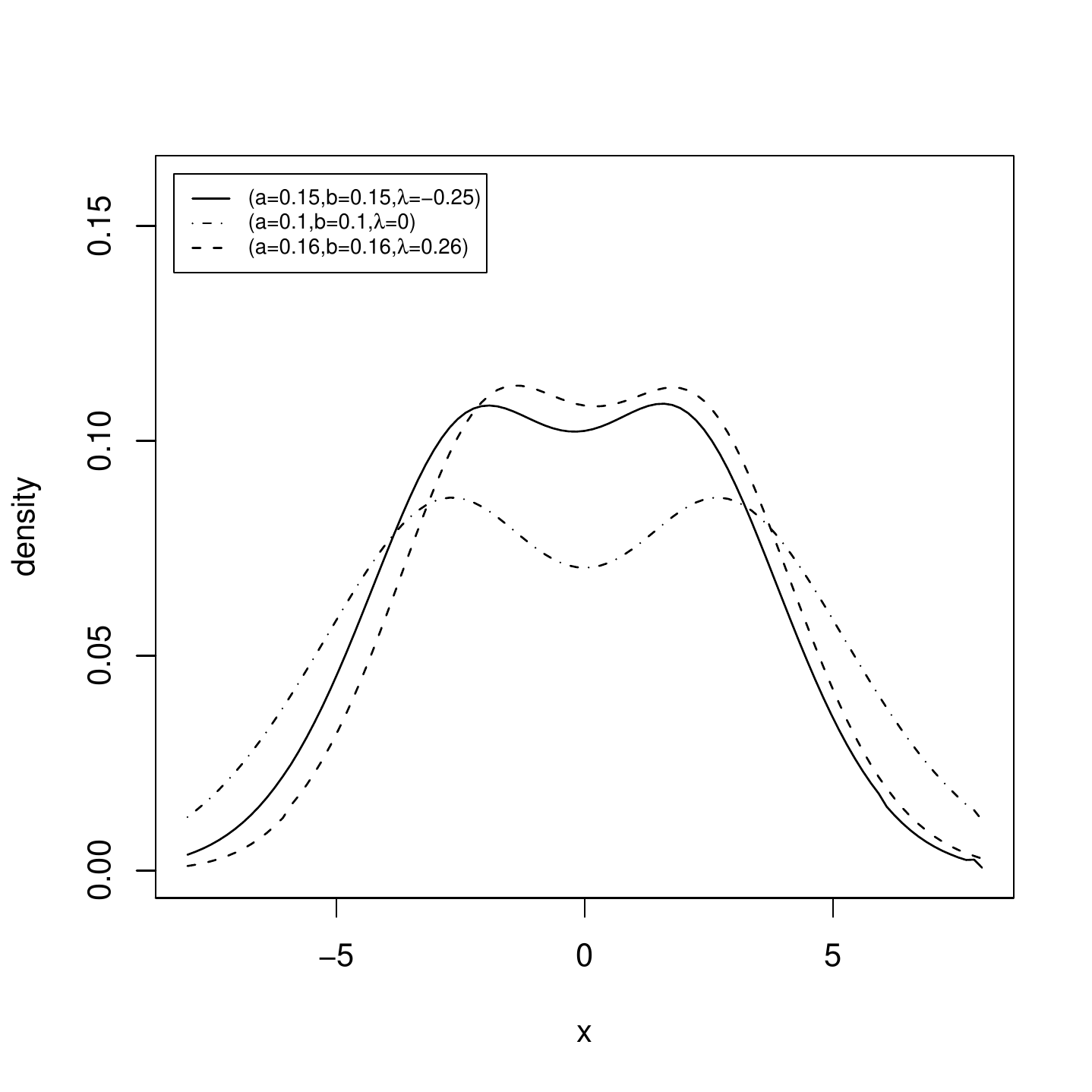}
\caption{\textit{The Beta skew-normal $BSN(\lambda,a,b)$ for values of $a,b <1$}}
\end{figure}
\end{document}